\documentclass[11pt,twocolumn]{article}

\usepackage[margin=1in]{geometry}  
\usepackage{graphicx}              
\usepackage{amsmath}               
\usepackage{mathtools, cuted}
\usepackage{amsfonts}             
\usepackage{amsthm}                
\usepackage{setspace}  
\usepackage{lineno} 
\usepackage{indent first}
\usepackage{balance}
\usepackage{gensymb}
\usepackage{titlesec}
\usepackage{titling}
\usepackage{booktabs}
\usepackage{multicol}
\usepackage{etoolbox}
\usepackage{threeparttable}
\usepackage{appendix}
\usepackage{authblk}
\usepackage[round, numbers]{natbib}
\usepackage{newfloat}

\usepackage[font={footnotesize,singlespacing},labelfont=bf]{caption}
\DeclareFloatingEnvironment[name={Figure S}]{appfig}
\DeclareFloatingEnvironment[name={Table S}]{apptab}
\bibpunct{(}{)}{,}{a}{}{,}

\newcommand{\eqname}[1]{\tag*{#1}}

{\large }



\newbool{MyRefNumbers}
\booltrue{MyRefNumbers} 
\titlespacing*{\section}
{0pt}{2ex plus 1ex minus .2ex}{2ex plus .2ex}
\titlespacing*{\subsection}
{0pt}{2ex plus 1ex minus .2ex}{2ex plus .2ex}

\titleformat*{\subsection}{\normalsize \itshape}

\usepackage{fancyhdr}
\begin{document}
\onecolumn
\title{Stochastic density effects on adult fish survival and implications for population fluctuations}
\author[1]{Daniel K. Okamoto\thanks{corresponding author: dokamoto@sfu.ca }\thanks{Present address: School of Resource and Environmental Management/Hakai Network\\ Simon Fraser University, Burnaby, BC V5A 1S6, Canada
}}
\author[1,2]{Russell J. Schmitt\thanks{email: schmitt@lifesci.ucsb.edu}}
\author[1,2]{Sally J. Holbrook\thanks{email: sally.holbrook@lifesci.ucsb.edu}}
\affil[1]{Department of Ecology, Evolution and Marine Biology
 \\University of California Santa Barbara\\Santa Barbara, CA 93106 USA}
\affil[2]{Marine Science Institute\\ University of California Santa Barbara\\Santa Barbara, CA 93106 USA }
\renewcommand\Authands{ and }
\date{Updated January 22, 2016}
\maketitle
\leavevmode\kern15pt
\begin{flushleft}

\textbf{Citation}:  Okamoto, D. K., Schmitt, R. J. and Holbrook, S. J. (2016), Stochastic density effects on adult fish survival and implications for population fluctuations. \emph{Ecology Letters}, 19(2): 153?162. doi:10.1111/ele.12547. 

\bigskip

\end{flushleft}

\section*{Abstract}
The degree to which population fluctuations arise from variable adult survival relative to variable recruitment has been debated widely for marine organisms.  Disentangling these effects remains challenging because data generally are not sufficient to evaluate if and how adult survival rates are regulated by stochasticity and/or population density.  Using unique time-series for a largely unexploited reef fish, we found both population density and stochastic food supply impacted adult survival.  The estimated effect of variable survival on adult abundance (both mean and variability) rivaled that of variable recruitment.  Moreover, we show density dependent adult survival can dampen impacts of stochastic recruitment.  Thus, food variability may alter population fluctuations by simultaneously regulating recruitment and compensatory adult survival.  These results provide an additional mechanism for why intensified density independent mortality (via harvest or other means) amplifies population fluctuations and emphasizes need for research evaluating the causes and consequences of variability in adult survival.
\twocolumn
\section*{Introduction}
Populations can exhibit complex patterns of temporal variability when vital rates react to both extrinsic forcing and density dependent regulation \citep{Bjornstad2001}.  As shown for fish populations among others, one consequence of such an interaction is that intensification of adult mortality can increase variability in abundance or biomass \citep{Rouyer2012, Hsieh2006}.  Mechanistic hypotheses explaining this phenomenon include cohort resonance \citep{Botsford2014, Worden2010, Bjornstad2004}, increased environmental tracking via age truncation \citep{Hsieh2006} and increased intrinsic instability \citep{Shelton2011, Anderson2008}.  These and other complex population responses may be exacerbated if adult survival rates also exhibit stochasticity and density dependence.  While these processes are expected to exert less of an impact on survival of reproductive adults than recruitment \citep{Eberhardt2002, Pfister1998}, canalization of adult survival is not a ubiquitous expectation.  Survival of adults in some species reacts strongly to environmental stochasticity \citep{Grosbois2008} and in others exhibits density dependence [e.g., rodents \citep{Leirs1997}, birds \citep{Barbraud2003} and ungulates \citep{Bonenfant2009}].  Moreover, many key findings of models caution against ignoring variation in survival of reproductive adults.  These include higher elasticity of growth rates to adult survival than recruitment processes \citep{Gaillard2003, Heppell2000}, dampening or amplification of variance in adult population biomass when adult survival rates covary negatively or positively with recruitment \citep{Shelton2011}, similar effects of stochastic survival of different ages in models of salmon \citep{Worden2010} and hydra effects resulting from density dependent vital rates \citep{Abrams2009}.  Thus, adult survival that is subject to both density dependence and stochasticity can have important consequences for population size and variability.

In many systems, and for marine species in particular, researchers focus on the mean adult survival rather than its variability.  Simplifying assumptions of static adult survival (i.e., independent of density or stochasticity) arise out of necessity when existing data and knowledge cannot describe more complicated processes.  Fisheries stock-assessment models, for example, typically assume constant survival, which in many cases is poorly estimated or simply chosen \citep{Quinn1999}.  Despite this general convention, known pitfalls of using erroneous fixed estimates of adult survival include inaccurate assessments of productivity, temporal trends in abundance, or resilience to perturbation \citep{Johnson2015}.  Two key factors, among others, that are likely to shape variability in adult survival are density dependence and stochastic food variability.

Many challenges hinder understanding population-level implications of adult food limitation.  In competitive systems, resource limitation can arise via changes in resource availability and/or changes in adult density, and both can lead to reductions in fecundity as well as in survivorship of juveniles and adults \citep{Eberhardt2002, Mduma1999, Elliott1998, Clutton-Brock1997}.  Adults of a variety of vertebrates can buffer against food limitation by adjusting fecundity \citep{Clutton-Brock1997} or skipping reproduction \citep{Rideout2011}.  For fish, however, sacrificing fecundity first under nutritional stress is neither expected nor observed to be a universal trait \citep{Rideout2011} and adults doing so are likely to experience decreases in survival once reproductive energy stores are expended.  Given the widespread evidence that adults of fish (and other vertebrates) occasionally experience nutritional stress and competition for food, these forces have potential to influence population dynamics in complex ways through adult survival.  Furthermore, effects of food supply are likely to differ at different densities \citep{Eberhardt2002}.  At lower densities, food supply may have little impact and recruitment dynamics may dominate population trends.  In contrast, at high adult densities decreased survival due to competition may dampen effects of variation in recruitment.

While examples of density dependent population regulation are widespread for many vertebrates, strong empirical evidence for density dependent survival of adults is generally less prevalent for marine fish.  Among marine reef fishes, there is considerable evidence for density dependent juvenile survival for species that shelter from predators \citep{Holbrook2002, Osenberg2002, Schmitt1999}.  For exploited fishes, stochasticity and density dependence in recruitment are widely argued to drive increased variability in the face of reduced survival due to fishing \citep{Botsford2014, Rouyer2012, Shelton2011, Worden2010, Bjornstad2004}.  If adults also exhibit density dependent survival, then the dynamics of adult populations, both in terms of steady states and patterns of variability, are likely to differ substantially from expectations under static survival.  In addition, reductions in mean survival should relax any density dependence and increase sensitivity of adult abundance to variable recruitment.

To our knowledge no studies have evaluated the individual and combined effects of food and density-mediated adult survival on dynamics of fish populations.  While it is both intuitive and generally accepted that such effects can affect population fluctuations in marine species, ecological data are generally insufficient to 1) evaluate their individual and combined impacts on survival and 2) estimate the effect such impacts are expected to exert on population dynamics.  In addressing this gap, we first use a case study of the black surfperch (\emph{Embiotoca jacksoni}) to weigh evidence for food mediated and density dependent adult survival.  Second, we use a generalized simulation across a range of life history characteristics to evaluate how density dependent adult survival can influence population variability and how that may interact with persistent changes in survival rates (such as those due to harvest, climatic shifts, etc.).  For the case study, we used a Bayesian state-space modeling approach to evaluate whether conspecific density and food supply simultaneously affect adult survival rates in black surfperch, and whether incorporating these effects alter expectations of adult population size and variance.

Black surfperch are demersal reef fish of the eastern North Pacific and provide a tractable system for studying this issue because they exhibit strong competition for foraging habitat (benthic turf and macroalgae) and the crustacean prey therein \citep{Holbrook1986, Holbrook1984, Schmitt1986, Schmitt1984, Hixon1981}.  Recruitment dynamics are correlated with variation in food supply \citep{Okamoto2012} with an average observed juvenile:adult ratio around 1:4.  We used multi-generational time series data for black surfperch from Santa Cruz Island, California to evaluate the evidence for, and expected implications of, food limited and density dependent adult survival.  First we tested whether estimated survival varied through time.  We then estimated support for models that relate survival to conspecific density, food supply and availability of foraging habitat.  We used these models to calculate effect size measures in terms of equilibrium densities and variation in population density.  Finally, under a more general framework we simulated the response of hypothetical populations to variability in recruitment under a factorial gradient of density dependent adult survival, long-term mean survival rates, mean recruitment productivity, and stock-recruit (S-R) relationships.  Our findings reveal an additional mechanism by which persistent decreases in adult survival (due to fishing, climatic shifts, etc.) can further magnify the amplitude of population variability.

\section*{Methods}
\subsection*{Field surveys}
Data on stage-specific abundance of black surfperch, the amount of foraging habitat and the availability of their food were collected at 11 fixed locations within 4 sites on the north shore of Santa Cruz Island, CA in autumn intermittently from 1982-1992 and annually from 1993-2009.  At each location, fixed 40x2m transects at 3, 6, and 9 m depth contours (the typical black surfperch depth range) were surveyed for black surfperch and their foraging habitat by divers using SCUBA, and samples were collected to estimate their principal crustacean food items.  Counts of fish distinguished among young-of-year, juveniles (1 year old) and adults ($\geq$ 2 years old) (Figure 1).

\begin{figure*}[t]
\includegraphics[width= 170mm]{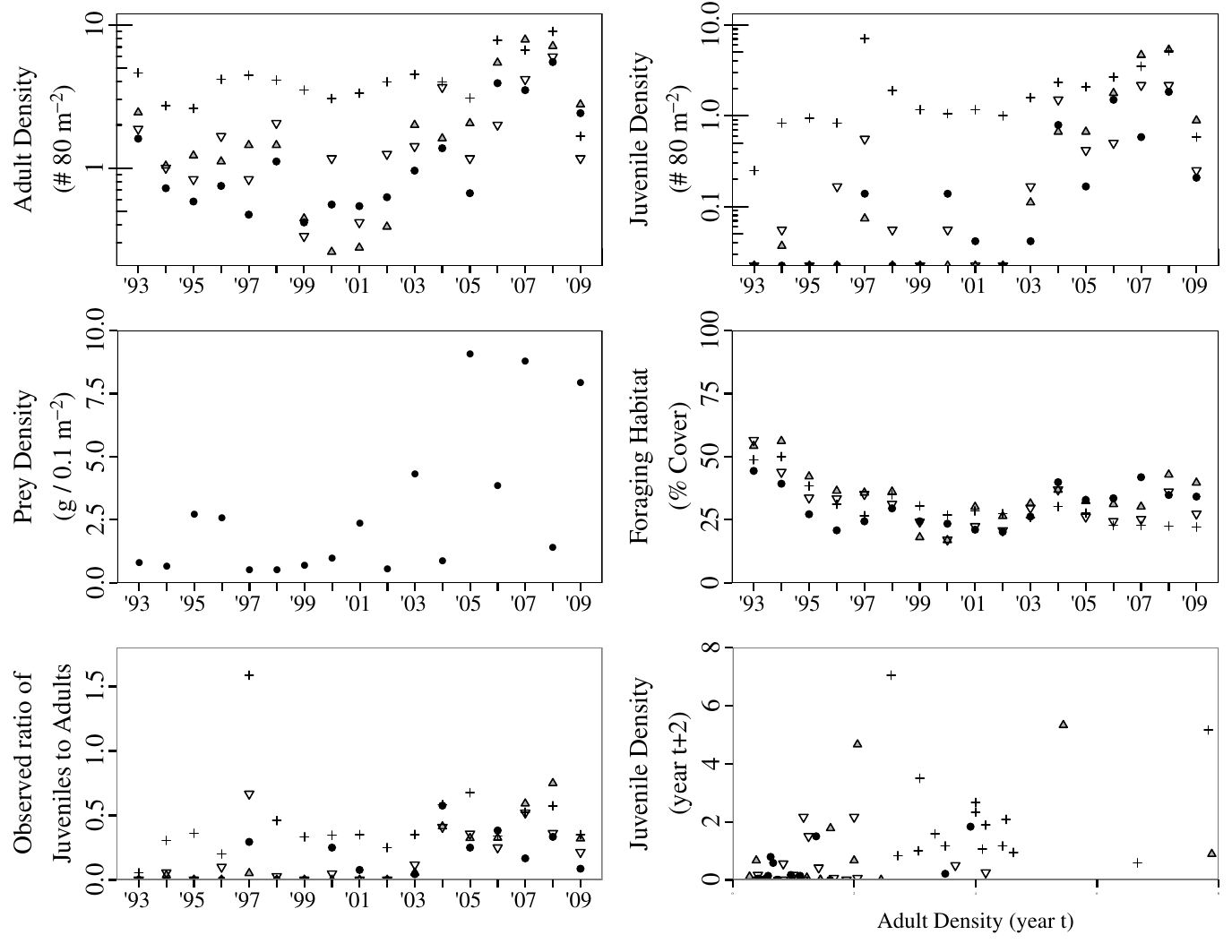}
\caption{A-D: Time series used in the study including adult density (A), juvenile density (B), prey density (C), and cover of foraging habitat (D).   E-F: demographic data including, observed juvenile:adult ratio at each site through time (E) and relationships between adults and subsequent juveniles two years later (F). Symbols segregate the four main sites on Santa Cruz Island where unique estimates exist. }  
\end{figure*}
\subsection*{Habitat and food density}
We defined foraging habitat as the average annual percent cover of all low-lying turf and foliose algae from which black surfperch harvest prey \citep{Laur1983, Holbrook1984, Schmitt1984} across all transects within each site.  Food density was defined as the observed biomass density of prey (g wet mass m-2), which included jaeropsid and idoteid isopods, gammarid and caprellid amphipods and crabs within the adult black surfperch gape limitation  \citep[Figure 1][]{Schmitt1984}.  See Appendix A for details.  Though predator densities also influence surfperch survival, regional predator abundance (primarily harbor seals and California sea lions) remained relatively consistent over the study period (Carretta et al. 2011).

\subsection*{Modeling framework and multimodel inference}
We used stage-structured, state-space models in a hierarchical Bayesian, multimodel framework to weigh evidence for relationships between survival rates and conspecific density, food availability and availability of foraging habitat.  Generating inference about density dependence from survey data presents distinct challenges \citep{Freckleton2006} because densities are observed with error and the underlying processes are unknown (the ``inverse problem").  Our framework allows us to incorporate measurement error, model uncertainty and associated parameter uncertainty by considering both observations and processes as latent unobservables and marginalizing inference over the range of plausible state variables, model structures and parameter values that may have generated the observations.  Bayesian inference approaches are commonly used in tackling inverse problems \citep{Mosegaard2002} and are capable of addressing the fundamental statistical issues they present \citep{Freckleton2006}, although the correlation versus causation issues remain unresolvable. 

In the stage structured model survival is a function of adult density in year ($y$) at the $i$th site ($A_{y,i}$), adults in the following year ($A_{y+1,i}$) and one-year-old (juvenile) density ($J_{y,i}$).  However, $A$ and $J$ were estimated from visual surveys and not truly known. Thus we used a state-space model formulated as follows:

\begin{align}
{A}_{y+1,i}& = [{A}_{y,i} {s}_{y,i}+{J}_{y,i}{s}_{y,i}g]e^{{\epsilon}_{y,i}}\label{eq:2.1} \\ 
{\hat{A}}_{y,i}& = {A}_{y,i} +{\epsilon}_{A,y,i}\label{eq:2.2}
\end{align}
where $s_{y,i}$ is adult survival from year ($y$) to ($y+1$), g is the maturation rate of juveniles in ($y$) to adulthood in ($y+1$) (juveniles either graduate or die), and $\hat{A}_y,i$ and$\hat{J}_y,i$ are the observed survey densities.  $\epsilon_{A_y,i}$ is the normal adult observation error (with variance $\sigma_{A}$) and $\epsilon_{y,i}$ is the lognormal process error (see Appendix A for the associated likelihoods).

There is no process model for juveniles or adults in ($y=1$). Here we incorporate observation error by imposing the following priors:
 
\begin{align}
 {J}_{y,i}&\sim \text{Normal}(\hat{J}_{y,i},\sigma_{J}^2) \label{eq:2.3}\\
 {A}_{y=1,i}&\sim \text{Normal}(\hat{A}_{y=1,i},\sigma_{A}^2) \label{eq:2.3b}
 \end{align}
where $\hat{J}_{y,i}$ is the observed value and $\sigma_J$ is the observation standard deviation.  We independently estimated $\sigma_A$ and $\sigma_J$ with an external training dataset (see Appendix B, Figure B.1).

We first evaluated evidence that survival varies by year using the product space method \citep{Lodewyckx2011, Carlin1995}.  We compared model support using Bayes factors \citep{Gelman2013, Kass1995}, which measure data-driven evidence in favor of a hypothesis after accounting for its prior probability, and 2 ln(Bayes factor) (hereafter referred to as 2lnB) transforms it to an interpretable scale.  Values of 2lnB $>10$ are very strong evidence against the null (requiring greater than 99\% support) used here as a conservative threshold, while negative values support the null.  We then estimated models that consider annual survival as a function of food availability ($P_{y,i}$) in ($y$) and ($y+1$), foraging habitat ($H_y$,) in ($y$) and ($y+1$), and conspecific density (adults and juveniles independently) with two model forms: the Logistic and the Shepherd.  The Logistic Model [eqn (5)] relates survival ($s_{y,i}$) to density ($A_{y,i}$ \& $J_{y,i}$), food ($P_{y,i}$) and availability of foraging habitat ($H_{y,i}$) using a logistic function:

\begin{align*}	
{ s }_{ y,i }&=\frac {1}{\splitfrac{1+\exp{[({ \beta  }_{ { 0 }_{ i } }+{ \beta  }_{ 1 }A_{ y,i }+{ \beta  }_{ 2 }{ J_{y,i}+\beta_{3 }P_{ y,i }+ }}}{\beta_{4 }P_{ y+1,i }+\beta_{5}H_{y,i }+\beta_{6}H_{ y+1,i } )t]}} \label{eq:2.4}\\[5pt]
\eqname{Logistic Model (5) }
\end{align*} 
where ${ \beta  }_{ { 0 }_{ i } }$ is a site-specific scale parameter, ${ \beta  }_{ 1 }-{ \beta  }_{ 6 }$ are coefficients and $t$ is the fraction of the year elapsed since the previous time period.

The Shepherd Model \citep{Shepherd1982} provides additional flexibility and ease of biological interpretation [as derived in \cite{Quinn1999}]:

\begin{align*}	
{ s }_{ y,i }&=\frac { e ^{ -{ z }_{ i }t }}{ 1+\left( 1-{ e }^{ -{ z }_{ i }t } \right) K \left({ { A }_{ y,i }}\right)^\gamma  }\label{eq:2.5a}    \\[10pt]
\eqname{Shepherd Model (6a)}
\end{align*}
where $e^{-z}$ is the site-specific density independent survival rate, $\gamma$ controls the intensity (shape) of density dependence, and K controls the strength of density dependence.  If one considers K to be a function of environmental covariates, then the Shepherd Model can be represented as a linear combination of log scale predictors and coefficients as shown in [eqn (6b)].

\begin{align*}	
{ s }_{ y,i }&=\frac{e ^{ -{ z }_{ i }t }}{
\splitfrac{\splitfrac{\splitfrac{ 1+\left( 1-{ e }^{ -{ z }_{ i }t }\right)  \exp \textbf{[}({ \beta  }_{ { 0 }_{ i } }+}
{ \gamma \ln{A _{ y,i }}+{ \beta  }_{ 2 }\ln{J_{y,i}}+ }}{\beta_{3 }\ln{P}_{ y,i }+\beta_{4 }\ln{P}_{ y+1,i }+}}{\beta_{5}\ln{H}_{y,i }+\beta_{6}\ln{H}_{ y+1,i } )t \textbf{]}}} 
\label{eq:2.5b} \\[10pt]
\eqname{(6b)}
\end{align*}	
The Shepherd Model provides unique flexibility because it can range from density independent ($\gamma \rightarrow 0$), to a decelerating but non-saturating density dependent form (0$<\gamma<1$), to a saturating form ($\gamma=1$), and finally to an overcompensatory form ($\gamma>1$) in which survivors (eventually) approach zero as abundance increases indefinitely.  However, this model includes an additional parameter because it separates the density independent survival rate ($e^{-z}$) from $\beta_0$ (a nuisance parameter).  Thus we used the Logistic Model [eqn (5)] for statistical simplicity and the Shepherd Model [eqn (6b)] for biological flexibility and compared emergent properties of the two for qualitative agreement.
 
We applied stochastic search variable selection [SSVS, \cite{George1993}] to these models using Gibbs sampling to evaluate which combination of variables exhibited strong correlations with adult survival rates.  SSVS searches across the multitude of unique covariate combinations and estimates the probability that each (including the null) should be included.  Following SSVS model selection we sampled posteriors of the best Shepherd and Logistic Model for analysis including only covariates supported by 2lnB$>10$ in the SSVS procedure.  We included an AR(1) model on the process error terms to account for potential bias due to serial autocorrelation.  In all models we used vague priors.  Appendix B includes posterior predictive checks \citep{Gelman2013}, prior specification and prior-posterior comparisons. 

 We conducted all analyses using R \citep{Rsoftware} and JAGS \citep{Plummer2013}.

\subsection*{Comparative effect size of food supply on equilibrium abundance}
Estimation of effect size is critical in generating inference (Osenberg et al. 2002). First we sought to determine the magnitude by which food supply and density dependence impact equilibrium abundance.  To quantify individual and combined effects of each significant model covariates, we combined correlative stock recruit models \citep{Okamoto2012} with the Shepherd Model [eqn (6b)] posteriors and solved for equilibrium density under steady food conditions.  We numerically estimated the equilibrium under a factorial gradient of food supply for adults and recruitment which allowed us to compare effect sizes of food supply on expected equilibria via adult survival and via recruitment under hypothetical steady conditions.

 \subsection*{Comparative effect size of adult density dependence and food supply on population variance}
 \begin{table*}[!ht]
\caption{Table of log-scale Bayes factors (2lnB) for variables influencing annual adult survival. Values greater than +10 are very strong evidence in favor of inclusion (shown in bold, requiring $>99\%$ support given even prior odds); negative values indicate the null has more support than the alternative hypothesis.  The values are calculated as 2lnB measuring strength of inference from the data in favor of the alternative hypothesis against the null.  See Methods for model descriptions. }
\begin{center}
\begin{tabular}{lcc}
  \toprule
Variable & Shepherd Model & Logistic Model \\ 
  \midrule
\textbf{Adult Density ($y$)} & \textbf{10.6} & \textbf{15.4} \\ 
  \textbf{Prey Density ($y$)} & \textbf{11.3} & \textbf{21.0} \\ 
  Juvenile Density  ($y$) & -4.3 & -6.0 \\ 
  Prey Density  ($y+1$) & -2.0 & -6.7 \\ 
  Foraging Habitat Availability ($y$) & -2.7 & -7.1 \\ 
  Foraging Habitat Availability ($y+1$) & -4.8 & -7.7 \\ 
   \bottomrule
\end{tabular}
\end{center}
\end{table*}
To assess the comparative effects of density dependent adult survival, recruitment variability, and food supply for adults on population variability, we simulated temporal variability in food supply and projected the population size through time (using the parameterized model from the previous section) under 4 different alternative scenarios: (1) recruitment is unchanged by food supply, (2) adult survival is compensatory but unchanged by food supply, (3) adult survival is constant [fixed at the mean of (2)] with no compensatory response, and (4) both recruitment and adult survival respond to food supply with compensatory adult survival.  We simulated temporal variability in food supply using a multivariate normal distribution (truncated at the observed range and the adult-recruit food supply correlation set to the observed value, r = 0.88).  Simulations included 1000 years with parameters from the posterior mean of the Shepherd Model and fitted values from the stock recruit dynamics from \citep{Okamoto2012}.  We simulated an autocorrelated food supply using Markov chain Monte Carlo tuned to provide a lag 1 partial autocorrelation (pacf) of $\phi$=0.65 (the observed value for recruitment of age 1 individuals at the site with the greatest age 1 abundance); $\phi$=0 provided qualitatively similar results.

\subsection*{Generalized buffering effect of density dependent adult mortality against recruitment variation}
 
While the theoretical population-level effect of density independent stochastic survival is well studied for an array of life history characteristics \citep{Shelton2011, Worden2010}, the general role that density dependent adult survival plays in altering population sensitivity to variability remains far less explored.  Specifically the sensitivity of adult fluctuations to recruitment variability should vary with key traits, including the strength of adult density dependence, the reproductive productivity, mean adult longevity, and strength of recruitment compensation.  To generalize the model into a flexible delay difference model we combined the basic Shepherd Model of adult survival [eqn (6a)] with a standard Cushing SR function.
 \begin{align*}
{A}_{y+1} & = \text{surviving adults}+\text{recruitment} \nonumber \\
=&\frac { { { { A }_{ y}}{ e }^{ -{ z }t } } }{ 1+\left( 1-{ e }^{ -{ z }t } \right) K \left({ { A }_{ y }}\right)^\gamma  }+\bar{\alpha}\left({ { A }_{ y-2 }}\right)^\beta e^{\epsilon_y- \frac{\sigma^2}{2}} \label{eq:2.6} 
\eqname{(7)}
\end{align*}
where $\alpha$ controls density independent per capita productivity, $\beta$ controls recruitment compensation, and the error $\epsilon_y$ has variance $\sigma^2$.  Note that juvenile survival to adulthood is now combined with the stock-recruit function.

\begin{figure*}[t]
\begin{center}
\includegraphics[width= 110mm]{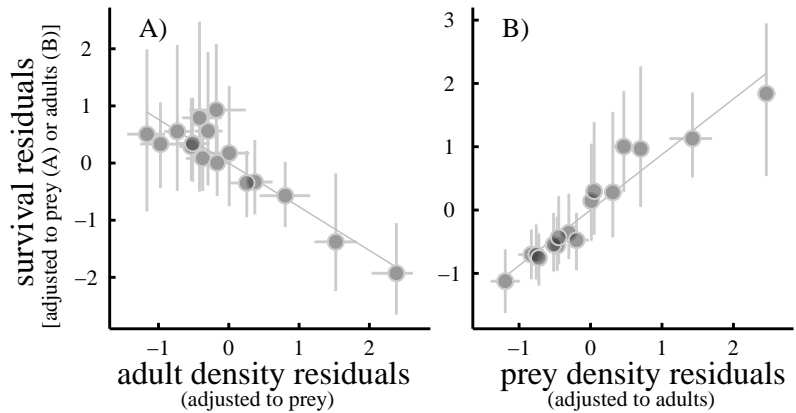}
\end{center}
\caption[Added variable plots]{Added variable plots generated using posterior samples from the Logistic Model showing isolated relationships between logit-scale survival and A) adult density and B) prey density in year ($y$).  Error bars show the upper and lower $95\%$ quantiles from the model posterior, averaged across sites.   Variables are adjusted to yield residuals by regressing both survival posterior estimates and the shown covariate against other covariates.  The line represents the slope of the mean estimated relationship in the Logistic Model where this visualization technique can be applied; because the nonlinear survival function of food and adult density in the Shepherd Model is not linear under direct transformation, similar plots are not shown for this model.  The scales represent standardized normal residuals. }  
\label{figure:F2.1}
\end{figure*}

Using this generalized model, we simulated how variability of the adult population responded to a gradient of temporal variability in $\alpha$ (controlled by $\epsilon$ via $\alpha=\bar{\alpha}e^{\epsilon_y-\frac{\sigma^2}{2}}, \mathrm{CV}[\alpha]=0.1-1)$, mean recruitment productivity ($\bar{\alpha}$), adult density dependence ($\gamma$), recruitment compensation ($\beta$), and long-term mean adult survival rate (controlled by adjusting z in each simulation) calculated as mean survival ($\bar{\text{s}}$) including density dependent and density independent processes.  $\gamma$ ranged from 0 (no density dependence) to 1 (saturating survival) up to 4 (strong overcompensation); $\bar{\alpha}$ from 0.25 to 1.50; $\bar{\text{s}}$ =0.5 (low), 0.65 (medium) and 0.8 (moderate) and ($\beta$)=0.25 (low) and 0.5 (moderate).  For each parameter combination we simulated 100 replicate 100 year time series.  See Appendix A for details.  We report results with pacf in $\alpha$ of $\phi$=0.65 ($\phi$=0 provided qualitatively similar results).  Because CV[adults] tended to increase linearly with CV[$\alpha$], we measured sensitivity of population fluctuations to recruitment variability as given by eqn (8).

\begin{align*}
\frac{\Delta {CV}_{adults}}{\Delta CV_{\alpha}}&\label{eq:2.7} \eqname{(8)}\end{align*}

For each scenario (combinations of $\bar{\alpha}$, $\gamma$, $\beta$ \&  $\bar{\text{s}}$) we quantified whether decreases in mean adult survival affect sensitivity to recruitment fluctuations by taking the difference in sensitivity across simulations with different mean survival rates.

\section*{Results}
Our analyses indicate survival rates of adult black surfperch varied among years with strong evidence against a single, static survival rate (2lnB=19.4).  The SSVS procedure with both the Logistic and Shepherd Models revealed strong correlations between annual survival and both food supply (positive) and adult conspecific density (negative) (Table 1; Figure 2).  The posterior parameter distribution suggests adult survival declined when density increased or food became scarce (Figure 2; Figure 3A).  Moreover, the impact of food supply diminished as adult density decreased, indicated by the similar survivorship across the range of food supply for low-density circumstances (Figure 3A).  Despite the weight of evidence in favor of interactions observed, there was also substantial uncertainty in the estimated effects of adult density and food supply.  The Shepherd Model indicates competition exists, with parameter estimates in the 95\% credible set ranging from very weak to strong overcompensatory dynamics (Figure 3A).  Likewise, effects of food are strong but highly uncertain where increased food supply and adult density occur together, illustrated by highly variable predictions of survivorship (Figure 3A).  The Logistic Model provided qualitatively similar results (see Appendix B, Figure B.3).  Posterior predictive checks suggest the models fit the data well and meet basic model assumptions (see Appendix B, Figure B.4).

\begin{figure}[!ht]
\begin{center}
\includegraphics[trim={1mm 1mm 1mm 1mm},clip,width= 75mm]{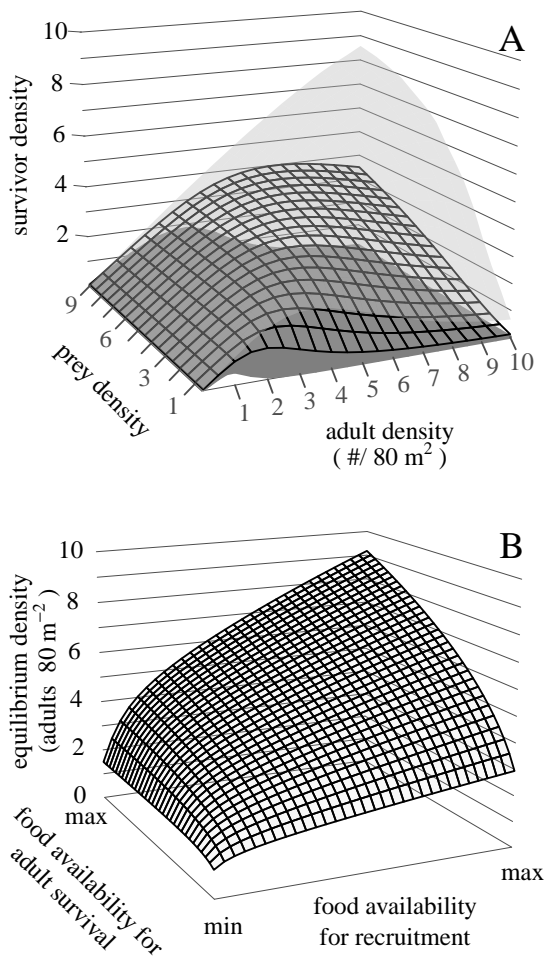}
\end{center}
\caption{A) Survivorship as a function of both adult surfperch density and density of food based on posterior samples from the modified Shepherd Model [eqn (6b)].  The gridded surface represents the posterior mean expectation, while the grey surfaces represent the 95\% posterior credible set for survival ($z$-axis) in each $x$-$y$ combination.  For a similar plot from the Logistic Model see Appendix B. Estimated mean effect size of food availability on equilibrium density via recruitment and via adult survival from the correlative survival and recruitment models.  The surface represents the prediction from the posterior mean from the Shepherd Model and the previously estimated recruitment models \citep{Okamoto2012}.  The equilibrium density was solved numerically for each intersecting line, assuming a constant environment therein.}  
\end{figure}

\subsection*{Effect size of food supply on equilibrium population size}
Estimates of adult equilibrium density are highly sensitive to food supply via both adult survival and recruitment across the range of observed food supply values.  Unsurprisingly, these effects also combine to increase equilibrium densities.  The impacts of food via recruitment and via adult survival, in this case, were estimated to be additive and of near equal importance (Figure 3B).  These estimated population size equilibria assume temporally constant environmental conditions.

\subsection*{Comparative effect size of nonlinearity and food supply on population variability}
Simulation analyses revealed three key findings with respect to how population variability responds to fluctuations in food supply.  First, the importance of simulated variation in recruitment on adult variability was diminished by compensatory adult survival (Figure 4).  Second, simulated population variability was impacted similarly by variation in both recruitment and adult survival due to variability in food supply (Figure 4).  Third, simulated variability in adult numbers responded more strongly to the combined effects of recruitment variation and variation in adult survival due to food than individual effects alone (Figure 4).  Variability in simulated adult density increased steeply with variability in food when variation in recruitment was driven by food supply and adult survival.  However, by incorporating adult density dependent survival, this variability in adult population size was minimized because not all recruitment variation is transferred to adults (Figure 4). The separate impacts of food variability via recruitment and density dependent adult survival are similar in magnitude and when combined markedly increased variability.

\subsection*{Generalized effects of density dependent adult mortality on population variance}
Adult compensatory survival buffers against stochastic recruitment variability.  This result is consistent across a factorial gradient of recruitment productivity ($\bar{\alpha}$), survival compensation  ($\gamma$) as well as multiple values of $\beta$ (recruitment compensation) and mean survival($\bar{\text{s}}$) (Figure 5 A-F).  The strongest dampening effects occur with overcompensatory adult survival ($\gamma>1$) but are also present with undersaturating ($\gamma=1$) and weak ($\gamma<1$) density dependence.  The impact of density dependence increases with $\bar{\text{s}}$. Moreover, this impact is present (though not equally strong) for all given values of $\bar{\text{s}}$ and $\beta$.  Thus, the dampening effect of density dependent adult survival occurs across a range of life history situations beyond the surfperch system, which exhibits intermediate mean annual survival rates ($\approx0.71\text{yr}^{-1}$) and modest recruitment compensation ($\beta \approx0.37$).

More importantly the strength of the dampening effect (contour steepness in Figure 5 A-F) decreases with mean survival ($\bar{\text{s}}$).  Thus decreases in s¯ increase sensitivity of adult population variability to recruitment fluctuations illustrated by the deviations in sensitivity between mean survival scenarios (Figure 5 G-J).  Those increases are generally more pronounced with greater density dependent adult survival.  This more pronounced change in sensitivity is not due to decreases in mean adult abundance, which stays constant as $\gamma$ increases for any given combination of mean survival ($\bar{\text{s}}$), density dependence ($\gamma$) and recruitment productivity ($\bar{\alpha}$).  The baseline change in sensitivity (when $\gamma$ = 0) occurs due to decreases in abundance.  Additional changes in sensitivity (when $\gamma>0$ in Figure 5 G-J) occur due to weakening of compensatory feedbacks.  The exception to this result occurs in the cases where mean adult survival is $0.8 \text{yr}^{-1}$ with high recruitment productivity and strong density dependent adult mortality (upper right portions of Figure 5 I \& J).  Here, changes in survival induce similar increases in sensitivity whether adult survival is density independent or strongly density dependent.  In all other regions density dependent adult survival exacerbates the amplification in temporal variability induced by reductions in mean survival.

\section*{Discussion}

\begin{figure}[ht]
\begin{center}
\includegraphics[width= 75mm]{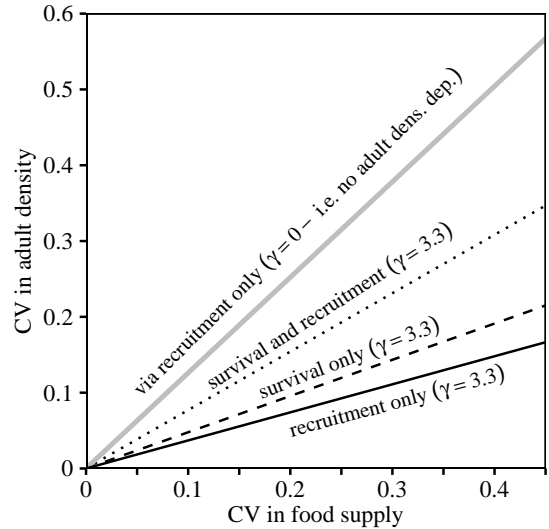}
\end{center}
\caption[Sensitivity of population variance]{Effect of food variability on fluctuations in adult density under different model assumptions shown as the coefficient of variation (CV) in adult density versus CV in food supply.  \textbf{Top (grey) line}: food variability only impacts the stock recruit function, survival is density independent and unaffected by food; \textbf{bottom (black) line}: food only impacts the stock recruit function, survival is density dependent but unaffected by food; \textbf{dashed line}: food variability impacts only adult survival (with density dependence) and the stock recruit function is constant; \textbf{dotted line}: food variability impacts both recruitment and density dependent adult survival.  \textbf{Top (grey) vs bottom (black) line}: Reduction in slope due to density dependent adult survival alone. \textbf{Top (grey) vs dashed line}:  Isolated effects of food via adult survival vs recruitment. See text for simulation details.}  
\end{figure}

Dramatic decreases in food supply should eventually reduce adult survival, especially when densities of competing adults are high.  Whether survival actually decreases should depend upon the magnitude of temporal food variability and the extent to which reductions in fecundity, energy reserves or somatic growth can buffer survival against nutritional stress in adults \citep{Gaillard2003}.  Here we provide strong evidence that variation in adult density and food abundance both alter survival rates of adult fish.  We show that the estimated relationships provide very different expectations of population equilibria and temporal variance in adult abundance compared to the case where survival is assumed to be constant (Figure 3; Figure 4).  For black surfperch the estimated effect size from food-driven variation in adult survival rivaled that of food-driven recruitment both in terms of estimated equilibrium density (Figure 3B) and variance (Figure 4).  While surfperch are in some ways a unique system, we show that density dependent adult survival is expected to dampen the impact of recruitment variation on adult population variability across a gradient of longevity, productivity and compensation (Figure 5).  This property facilitates increased temporal variance when the long-term mean survival rate decreases (Figure 5).  Thus while parameter estimates of the surfperch models include large uncertainty (Figure 3A) and likely are specific to this species, the general conclusions from these analyses apply across a spectrum of species characteristics (Figure 5).

\begin{figure*}[ht]
\begin{center}
\includegraphics[width= 170mm]{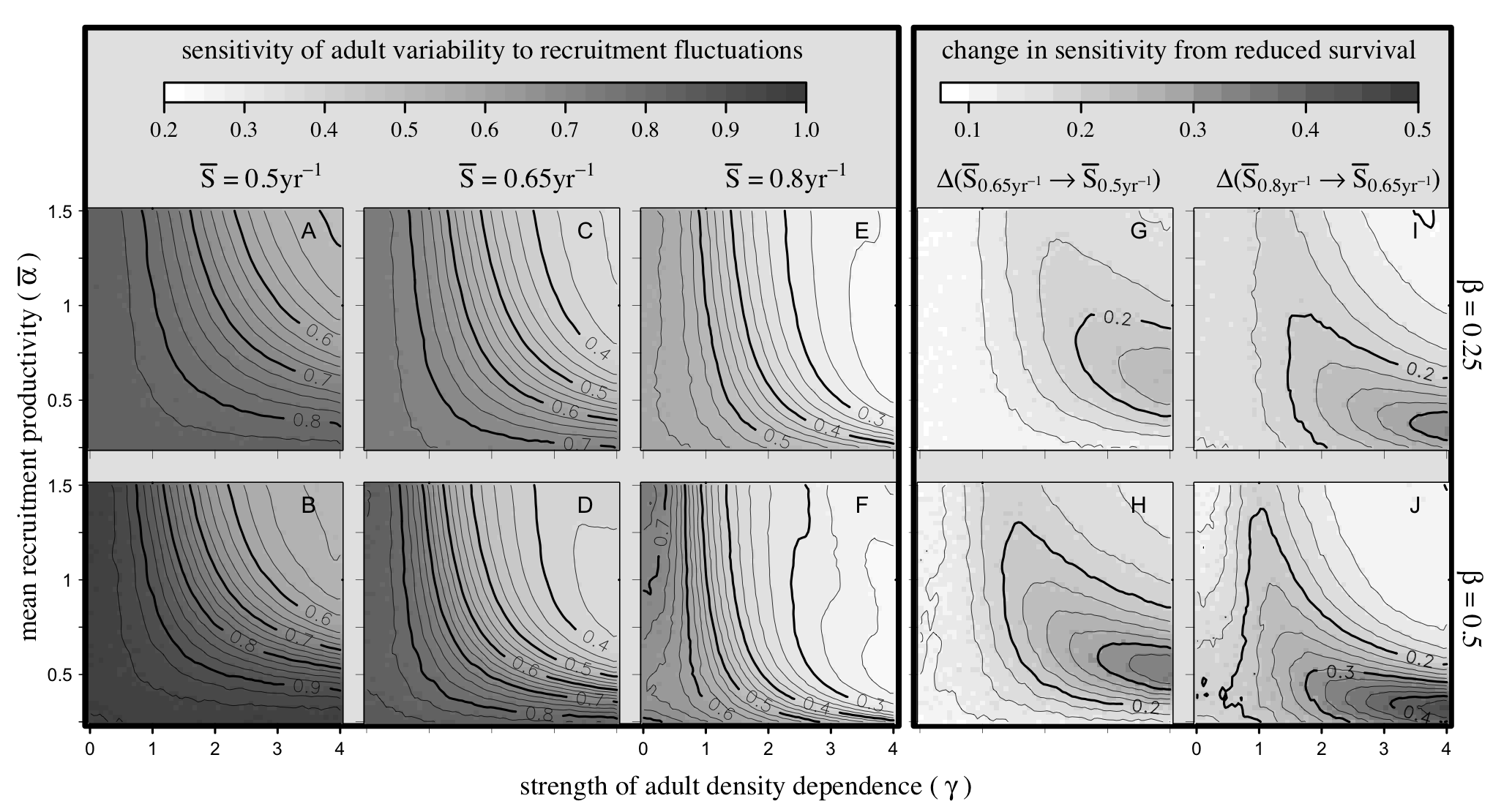}
\end{center}
\caption{Sensitivity of adult population variance to stochastic recruitment variability (A-F) and change in sensitivity due to reductions in survival (G-J) for simulations from the generalized population model [eqn (7)]. Axes represent adult density dependence ($\gamma$)  and mean recruitment productivity ($\bar{\alpha}$)  for two different recruitment compensatory strengths ($\beta$=0.25, 0.5).  Panels A-F represent three different mean annual adult survival rates ($\bar{\text{s}}$=0.5, 0.65, and 0.8yr$^{-1}$).  Panels A-F represent different mean adult survival rates (0.8 vs 0.65yr$^{-1}$ and 0.65 vs 0.5yr$^{-1}$). Stochasticity influences only $\alpha$ and for each simulation density independent adult mortality (z) is adjusted so the mean adult survival rate equals the desired value.  Sensitivity of adult variability is the slope of the relationship between CV in adult density and CV in $\alpha$ (analogous to the slope of lines in Figure 4). Simulations for each $\bar{\alpha}$-$\gamma$ combination included 100 replicate 100 year time series for CV($\alpha) \in$ 0.1, 0.2, $\dots$,1.0.}  
\end{figure*}

Several questions remain unanswered regarding why food and density appear to regulate survival of adult black surfperch.  The mechanism underlying density-survival interactions remains uncertain because we cannot disentangle processes that determine annual survival.  Our data contain no direct estimates of the demographics of adult mortality.  In addition, the estimated effects from food and density may arise from additional extrinsic factors such as predation or habitat availability.  Predation, for example, may provide the ultimate source of mortality simply because weak fish are eaten \citep{Cushing1975}.  Despite a steady predator abundance over the duration of the study \citep{Carretta2011}, per capita predation rates may vary as food for surfperch becomes scarce and they become weaker or spend more time foraging \citep{Holbrook1988}.  Similarly, declines in foraging habitat \citep{Holbrook1988, Holbrook1986, Holbrook1984, Schmitt1986, Hixon1981} may increase susceptibility of individuals to fluctuations in food density within those habitats.  Prior to the period covered by this study, a decrease in foraging habitat availability corresponded with a decline in black surfperch abundance \citep{Okamoto2012}.  Since then, foraging habitat availability varied little compared to variability in food density within those habitats.  Black surfperch may therefore require higher food density currently than when foraging habitat availability was greater.  These potential interactions highlight that the strength of density-resource interactions may change over time via several interacting mechanisms.  Thus, while there is a rich history of incorporating predator, food and ratio dependence into behavioral responses \citep{Abrams2000}, for most time series studies of population dynamics, the present included, the ability to capture the ``true" functional responses rather than just general correlations is an elusive challenge. 

This study highlights that even high quality time series data can yield low confidence in estimated demographic parameters.  For many species, even estimating a mean survival rate is challenging \citep{Johnson2015, Quinn1999}.  Precisely estimating annual vital rates often requires costly mark-recapture or other methods.  Our data, for example, were sufficient to infer the presence or absence of correlations and differences in survival among years, but parameter estimates include large uncertainty.  We illustrate, however, that under the posterior distribution, mean equilibrium adult population size and adult variance are strongly dependent on extrinsic and intrinsic factors related to adult survival.  Despite the known potential to mischaracterize dynamics of fish populations by assuming stationary adult survival \citep{Johnson2015}, the capacity to reliably estimate temporal variability in adult survival remains elusive.  We echo \cite{Shelton2011} regarding the importance of investing in research focused on the consequences of variability in survival as well as the underlying causes.

A key issue in ecology is how natural extrinsic and anthropogenic forces interact with intrinsic population properties to shape trends and variability in population size.  Both theory and observation suggest that reduced survival rates of adult fish can lead to greater recruitment variability \citep{Minto2008, Hsieh2006}.  These observations can arise from several mechanisms related to density dependent stock-rercruitment (S-R) relationships.  Cohort resonance \citep{Bjornstad2004} can arise from stable, stochastic S-R relationships \citep{Botsford2014, Rouyer2012, Worden2010} and is magnified by declines in mean adult abundance \citep{Botsford2014, Worden2010}.  Decreased adult survival may also lead to increased instability in the presence of overcompensatory S-R forms \citep{Shelton2011, Anderson2008}.  Finally, interannual ``bet-hedging" tactics of longer lived species \citep{Winemiller1992} are undermined by decreased survival due to truncation of adult age structure \citep{Anderson2008, Hsieh2006}, which diminishes inertia of adult populations.  The aforementioned studies explicitly evaluate the case where $\gamma$=0.  The observations from our study can exacerbate such effects in several ways.  First, simple stochasticity in adult survival can magnify the impact of increases in recruitment variability (Figure 4).  Second, our results show that density dependent adult survival dampens the sensitivity of variability in the adult population to stochastic recruitment (Figure 5 G-J).  For a given mean survival rate, the sensitivity of adult abundance to recruitment variability is inversely related to the strength of density dependent survival.  These results occur with and without overcompensatory dynamics (i.e. $\gamma >1$ and $\leq1$) that can generate instability.  Reductions in mean adult survival relax this compensatory effect and thereby increase sensitivity to recruitment variability.  This process exacerbates the relaxation in compensation expected when ($\gamma$=0).  However, for two reasons these compensatory effects are less plausible for species with very high mean annual adult survival.  First, variance in survival necessarily decreases as the mean approaches 1 $[\mathrm{Var}[\text{s}]\leq\bar{\text{s}}(1-\bar{\text{s}})]$.  Second, species with greater longevity and greater age at maturation are less sensitive to effects such as cohort resonance \citep{Botsford2014} and recruitment stochasticity in general \citep{Shelton2011}.  These taxa tend to fall in the spectrum of ``periodic" life-history strategists \citep{Winemiller1992} with low frequency recruitment success and high population inertia that buffer adult population size against recruitment volatility.  We speculate that compensation in adult survival is most likely in species with modest mean adult survival, early to modest maturation and strong competition for fluctuating resources.  In those cases reduced adult survival will tend to amplify transient trajectories of adult abundance.

Our findings provide important insight into the dynamics of stage and age structured populations.  When both adult survival and recruitment are controlled by resource availability, there is a potential two-fold impact.  Positive covariance between adult survival and recruitment can magnify the influence of the other \citep{Shelton2011}.  For black surfperch reducing prey availability leads to decreases in the number of recruits per adult \citep{Okamoto2012} and lowers the number of reproductive adults.  Such extrinsic influences can be buffered or amplified by intrinsic nonlinearities and cohort effects \citep{Lindstrom2002} and recent studies suggest decreases in overall survival via fishing or other persistent changes may increase population fluctuations \citep{Botsford2014, Rouyer2012, Shelton2011, Worden2010, Anderson2008}.  Here we illustrate an important mechanism that can exacerbate such effects: compensatory adult survivorship dampens fluctuations driven by recruitment variability.  Reductions in mean survival reduce this feedback and make populations more sensitive to stochastic recruitment.  Given these nuances, standard metrics of population resilience and stability may be dangerously misleading if density dependence and extrinsic factors individually or interactively drive survival of reproductive individuals.  Therefore long-term and process-based studies are needed to estimate how adult survival rates vary in time, what drives such variability, how adult survival covaries with recruitment, and what consequences ensue when adult survival is affected by interactions between extrinsic and intrinsic forces.

 \section*{Acknowledgements}
Funding for this research was provided by the National Science Foundation in support of the Santa Barbara Coastal Long Term Ecological Research (LTER) site (NSF OCE 1232779) and earlier awards to RJS and SJH from NSF and the US Minerals Management Service (now Bureau of Ocean Energy Management).  This work would not be possible without the valuable insight of Cherie Briggs, Robert Warner and Dan Reed.  The field and lab assistance from Keith Seydel, Clint Nelson, Shannon Harrer and Jessica Nielsen, as well as countless others who gathered the time series data are deeply appreciated.  We also thank Lou Botsford, Sergio Navarrete and three anonymous reviewers whose criticism substantially improved this work.

\singlespacing
\fontsize{11}{11} \selectfont
\bibliographystyle{ecol_let.bst}
\bibliography{Okamoto_et_al.bib}

\newcounter{MyBibCount}\providebool{MyRefNumbers}\begin{thebibliography}{50}
\expandafter\ifx\csname natexlab\endcsname\relax\def\natexlab#1{#1}\fi

\bibitem[{Abrams(2009)}]{Abrams2009}
\ifbool{MyRefNumbers}{\stepcounter{MyBibCount}\theMyBibCount.\\}{}Abrams, P.A.
  (2009).
\newblock When does greater mortality increase population size? The long
  history and diverse mechanisms underlying the hydra effect.
\newblock \emph{Ecology Letters}, 12, 462--474.

\bibitem[{Abrams \& Ginzburg(2000)}]{Abrams2000}
\ifbool{MyRefNumbers}{\stepcounter{MyBibCount}\theMyBibCount.\\}{}Abrams, P.A.
  \& Ginzburg, L.R. (2000).
\newblock The nature of predation: prey dependent, ratio dependent or neither?
\newblock \emph{Trends in Ecology \& Evolution}, 15, 337--341.

\bibitem[{Anderson \emph{et~al.}(2008)Anderson, Hsieh, Sandin, Hewitt,
  Hollowed, Beddington, May \& Sugihara}]{Anderson2008}
\ifbool{MyRefNumbers}{\stepcounter{MyBibCount}\theMyBibCount.\\}{}Anderson,
  C.N.K., Hsieh, C.h., Sandin, S.A., Hewitt, R., Hollowed, A., Beddington, J.
  \emph{et~al.} (2008).
\newblock Why fishing magnifies fluctuations in fish abundance.
\newblock \emph{Nature}, 452, 835--839.

\bibitem[{Barbraud \& Weimerskirch(2003)}]{Barbraud2003}
\ifbool{MyRefNumbers}{\stepcounter{MyBibCount}\theMyBibCount.\\}{}Barbraud, C.
  \& Weimerskirch, H. (2003).
\newblock Climate and density shape population dynamics of a marine top
  predator.
\newblock \emph{Proceedings of the Royal Society of London B: Biological
  Sciences}, 270, 2111--2116.

\bibitem[{Bj{\o}rnstad \& Grenfell(2001)}]{Bjornstad2001}
\ifbool{MyRefNumbers}{\stepcounter{MyBibCount}\theMyBibCount.\\}{}Bj{\o}rnstad,
  O.N. \& Grenfell, B.T. (2001).
\newblock Noisy clockwork: time series analysis of population fluctuations in
  animals.
\newblock \emph{Science}, 293, 638--643.

\bibitem[{Bj{\o}rnstad \emph{et~al.}(2004)Bj{\o}rnstad, Nisbet \&
  Fromentin}]{Bjornstad2004}
\ifbool{MyRefNumbers}{\stepcounter{MyBibCount}\theMyBibCount.\\}{}Bj{\o}rnstad,
  O.N., Nisbet, R.M. \& Fromentin, J.M. (2004).
\newblock Trends and cohort resonant effects in age-structured populations.
\newblock \emph{Journal of Animal Ecology}, 73, 1157--1167.

\bibitem[{Bonenfant \emph{et~al.}(2009)Bonenfant, Gaillard, Coulson,
  Festa‐Bianchet, Loison, Garel, Loe, Blanchard, Pettorelli \&
  Owen‐Smith}]{Bonenfant2009}
\ifbool{MyRefNumbers}{\stepcounter{MyBibCount}\theMyBibCount.\\}{}Bonenfant,
  C., Gaillard, J., Coulson, T., Festa‐Bianchet, M., Loison, A., Garel, M.
  \emph{et~al.} (2009).
\newblock Empirical evidence of density‐dependence in populations of large
  herbivores.
\newblock \emph{Advances in Ecological Research}, 41, 313--357.

\bibitem[{Botsford \emph{et~al.}(2014)Botsford, Holland, Field \&
  Hastings}]{Botsford2014}
\ifbool{MyRefNumbers}{\stepcounter{MyBibCount}\theMyBibCount.\\}{}Botsford,
  L.W., Holland, M.D., Field, J.C. \& Hastings, A. (2014).
\newblock Cohort resonance: a significant component of fluctuations in
  recruitment, egg production, and catch of fished populations.
\newblock \emph{ICES Journal of Marine Science: Journal du Conseil}, 71,
  2158--2170.

\bibitem[{Carlin \& Chib(1995)}]{Carlin1995}
\ifbool{MyRefNumbers}{\stepcounter{MyBibCount}\theMyBibCount.\\}{}Carlin, B.P.
  \& Chib, S. (1995).
\newblock Bayesian model choice via Markov chain Monte Carlo methods.
\newblock \emph{Journal of the Royal Statistical Society: Series B (Statistical
  Methodology)}, 57, 473--484.

\bibitem[{Carretta \emph{et~al.}(2011)Carretta, Forney, Oleson, Martien, Muto,
  Lowry, Barlow, Baker, Hanson, Lynch, Carswell, Brownell, Robbins, Mattila,
  Ralls, Opay, Norberg, Laake, Lawson, Cordaro, Petras, Sweetnam \&
  Yates}]{Carretta2011}
\ifbool{MyRefNumbers}{\stepcounter{MyBibCount}\theMyBibCount.\\}{}Carretta,
  J.V., Forney, K.A., Oleson, E., Martien, K., Muto, M.M., Lowry, M.S.
  \emph{et~al.} (2011).
\newblock U.S. Pacific Marine Mammal Stock Assessments.
\newblock {{U.S. Department of Commerce, NOAA Tech. Memo. NMFS-SWFSC-488}}.

\bibitem[{Clutton-Brock \emph{et~al.}(1997)Clutton-Brock, Illius, Wilson,
  Grenfell, MacColl \& Albon}]{Clutton-Brock1997}
\ifbool{MyRefNumbers}{\stepcounter{MyBibCount}\theMyBibCount.\\}{}Clutton-Brock,
  T.H., Illius, A.W., Wilson, K., Grenfell, B.T., MacColl, A.D.C. \& Albon,
  S.D. (1997).
\newblock Stability and instability in ungulate populations: an empirical
  analysis.
\newblock \emph{American Naturalist}, 149, 195--219.

\bibitem[{Cushing(1975)}]{Cushing1975}
\ifbool{MyRefNumbers}{\stepcounter{MyBibCount}\theMyBibCount.\\}{}Cushing, D.H.
  (1975).
\newblock The natural mortality of the plaice.
\newblock \emph{Journal du Conseil}, 36, 150--157.

\bibitem[{Eberhardt(2002)}]{Eberhardt2002}
\ifbool{MyRefNumbers}{\stepcounter{MyBibCount}\theMyBibCount.\\}{}Eberhardt,
  L.L. (2002).
\newblock A paradigm for population analysis of long-lived vertebrates.
\newblock \emph{Ecology}, 83, 2841--2854.

\bibitem[{Elliott \& Hurley(1998)}]{Elliott1998}
\ifbool{MyRefNumbers}{\stepcounter{MyBibCount}\theMyBibCount.\\}{}Elliott, J.M.
  \& Hurley, M.A. (1998).
\newblock Population regulation in adult, but not juvenile, resident trout
  (\emph{Salmo trutta}) in a Lake District stream.
\newblock \emph{Journal of Animal Ecology}, 67, 280--286.

\bibitem[{Freckleton \emph{et~al.}(2006)Freckleton, Watkinson, Green \&
  Sutherland}]{Freckleton2006}
\ifbool{MyRefNumbers}{\stepcounter{MyBibCount}\theMyBibCount.\\}{}Freckleton,
  R.P., Watkinson, A.R., Green, R.E. \& Sutherland, W.J. (2006).
\newblock Census error and the detection of density dependence.
\newblock \emph{Journal of Animal Ecology}, 75, 837--851.

\bibitem[{Gaillard \& Yoccoz(2003)}]{Gaillard2003}
\ifbool{MyRefNumbers}{\stepcounter{MyBibCount}\theMyBibCount.\\}{}Gaillard,
  J.M. \& Yoccoz, N.G. (2003).
\newblock Temporal variation in survival of mammals: a case of environmental
  canalization?
\newblock \emph{Ecology}, 84, 3294--3306.

\bibitem[{Gelman \emph{et~al.}(2013)Gelman, Carlin, Stern, Dunson, Vehtari \&
  Rubin}]{Gelman2013}
\ifbool{MyRefNumbers}{\stepcounter{MyBibCount}\theMyBibCount.\\}{}Gelman, A.,
  Carlin, J.B., Stern, H.S., Dunson, D.B., Vehtari, A. \& Rubin, D.B. (2013).
\newblock \emph{Bayesian Data Analysis}.
\newblock 3rd edn.
\newblock Chapman \& Hall, London.

\bibitem[{George \& McCulloch(1993)}]{George1993}
\ifbool{MyRefNumbers}{\stepcounter{MyBibCount}\theMyBibCount.\\}{}George, E.I.
  \& McCulloch, R.E. (1993).
\newblock Variable selection via Gibbs sampling.
\newblock \emph{Journal of the American Statistical Association}, 88, 881--889.

\bibitem[{Grosbois \emph{et~al.}(2008)Grosbois, Gimenez, Gaillard, Pradel,
  Barbraud, Clobert, M{\o}ller \& Weimerskirch}]{Grosbois2008}
\ifbool{MyRefNumbers}{\stepcounter{MyBibCount}\theMyBibCount.\\}{}Grosbois, V.,
  Gimenez, O., Gaillard, J., Pradel, R., Barbraud, C., Clobert, J.
  \emph{et~al.} (2008).
\newblock Assessing the impact of climate variation on survival in vertebrate
  populations.
\newblock \emph{Biological Reviews}, 83, 357--399.

\bibitem[{Heppell \emph{et~al.}(2000)Heppell, Caswell \& Crowder}]{Heppell2000}
\ifbool{MyRefNumbers}{\stepcounter{MyBibCount}\theMyBibCount.\\}{}Heppell,
  S.S., Caswell, H. \& Crowder, L.B. (2000).
\newblock Life histories and elasticity patterns: perturbation analysis for
  species with minimal demographic data.
\newblock \emph{Ecology}, 81, 654--665.

\bibitem[{Hixon(1981)}]{Hixon1981}
\ifbool{MyRefNumbers}{\stepcounter{MyBibCount}\theMyBibCount.\\}{}Hixon, M.A.
  (1981).
\newblock An experimental analysis of territoriality in the California reef
  fish \emph{Embiotoca jacksoni} (Embiotocidae).
\newblock \emph{Copeia}, 3, 653--665.

\bibitem[{Holbrook \& Schmitt(1984)}]{Holbrook1984}
\ifbool{MyRefNumbers}{\stepcounter{MyBibCount}\theMyBibCount.\\}{}Holbrook,
  S.J. \& Schmitt, R.J. (1984).
\newblock Experimental analyses of patch selection by foraging black surfperch
  (\emph{Embiotoca jacksoni} Agazzi).
\newblock \emph{Journal of Experimental Marine Biology and Ecology}, 79,
  39--64.

\bibitem[{Holbrook \& Schmitt(1986)}]{Holbrook1986}
\ifbool{MyRefNumbers}{\stepcounter{MyBibCount}\theMyBibCount.\\}{}Holbrook,
  S.J. \& Schmitt, R.J. (1986).
\newblock Food acquisition by competing surfperch on a patchy environmental
  gradient.
\newblock \emph{Environmental Biology of Fishes}, 16, 135--146.

\bibitem[{Holbrook \& Schmitt(1988)}]{Holbrook1988}
\ifbool{MyRefNumbers}{\stepcounter{MyBibCount}\theMyBibCount.\\}{}Holbrook,
  S.J. \& Schmitt, R.J. (1988).
\newblock Effects of predation risk on foraging behavior: mechanisms altering
  patch choice.
\newblock \emph{Journal of Experimental Marine Biology and Ecology}, 121,
  151--163.

\bibitem[{Holbrook \& Schmitt(2002)}]{Holbrook2002}
\ifbool{MyRefNumbers}{\stepcounter{MyBibCount}\theMyBibCount.\\}{}Holbrook,
  S.J. \& Schmitt, R.J. (2002).
\newblock Competition for shelter space causes density-dependent predation
  mortality in damselfishes.
\newblock \emph{Ecology}, 83, 2855--2868.

\bibitem[{Hsieh \emph{et~al.}(2006)Hsieh, Reiss, Hunter, Beddington, May \&
  Sugihara}]{Hsieh2006}
\ifbool{MyRefNumbers}{\stepcounter{MyBibCount}\theMyBibCount.\\}{}Hsieh, C.,
  Reiss, C.S., Hunter, J.R., Beddington, J.R., May, R.M. \& Sugihara, G.
  (2006).
\newblock Fishing elevates variability in the abundance of exploited species.
\newblock \emph{Nature}, 443, 859--862.

\bibitem[{Johnson \emph{et~al.}(2015)Johnson, Monnahan, McGilliard, Vert-pre,
  Anderson, Cunningham, Hurtado-Ferro, Licandeo, Muradian \& Ono}]{Johnson2015}
\ifbool{MyRefNumbers}{\stepcounter{MyBibCount}\theMyBibCount.\\}{}Johnson,
  K.F., Monnahan, C.C., McGilliard, C.R., Vert-pre, K.A., Anderson, S.C.,
  Cunningham, C.J. \emph{et~al.} (2015).
\newblock Time-varying natural mortality in fisheries stock assessment models:
  identifying a default approach.
\newblock \emph{ICES Journal of Marine Science: Journal du Conseil}, 72,
  137--150.

\bibitem[{Kass \& Raftery(1995)}]{Kass1995}
\ifbool{MyRefNumbers}{\stepcounter{MyBibCount}\theMyBibCount.\\}{}Kass, R.E. \&
  Raftery, A.E. (1995).
\newblock Bayes factors.
\newblock \emph{Journal of the American Statistical Association}, 90, 773--795.

\bibitem[{Laur \& Ebeling(1983)}]{Laur1983}
\ifbool{MyRefNumbers}{\stepcounter{MyBibCount}\theMyBibCount.\\}{}Laur, D.R. \&
  Ebeling, A.W. (1983).
\newblock Predator-prey relationships in surfperches.
\newblock \emph{Environmental Biology of Fishes}, 8, 217--229.

\bibitem[{Leirs \emph{et~al.}(1997)Leirs, Stenseth, Nichols, Hines, Verhagen \&
  Verheyen}]{Leirs1997}
\ifbool{MyRefNumbers}{\stepcounter{MyBibCount}\theMyBibCount.\\}{}Leirs, H.,
  Stenseth, N.C., Nichols, J.D., Hines, J.E., Verhagen, R. \& Verheyen, W.
  (1997).
\newblock Stochastic seasonality and nonlinear density-dependent factors
  regulate population size in an African rodent.
\newblock \emph{Nature}, 389, 176--180.

\bibitem[{Lindstr{\"o}m \& Kokko(2002)}]{Lindstrom2002}
\ifbool{MyRefNumbers}{\stepcounter{MyBibCount}\theMyBibCount.\\}{}Lindstr{\"o}m,
  J. \& Kokko, H. (2002).
\newblock Cohort effects and population dynamics.
\newblock \emph{Ecology Letters}, 5, 338--344.

\bibitem[{Lodewyckx \emph{et~al.}(2011)Lodewyckx, Kim, Lee, Tuerlinckx, Kuppens
  \& Wagenmakers}]{Lodewyckx2011}
\ifbool{MyRefNumbers}{\stepcounter{MyBibCount}\theMyBibCount.\\}{}Lodewyckx,
  T., Kim, W., Lee, M.D., Tuerlinckx, F., Kuppens, P. \& Wagenmakers, E.J.
  (2011).
\newblock A tutorial on Bayes factor estimation with the product space method.
\newblock \emph{Journal of Mathematical Psychology}, 55, 331--347.

\bibitem[{Mduma \emph{et~al.}(1999)Mduma, Sinclair \& Hilborn}]{Mduma1999}
\ifbool{MyRefNumbers}{\stepcounter{MyBibCount}\theMyBibCount.\\}{}Mduma,
  S.A.R., Sinclair, A.R.E. \& Hilborn, R. (1999).
\newblock Food regulates the Serengeti wildebeest: A 40 year record.
\newblock \emph{Journal of Animal Ecology}, 68, 1101--1122.

\bibitem[{Minto \emph{et~al.}(2008)Minto, Myers \& Blanchard}]{Minto2008}
\ifbool{MyRefNumbers}{\stepcounter{MyBibCount}\theMyBibCount.\\}{}Minto, C.,
  Myers, R.A. \& Blanchard, W. (2008).
\newblock Survival variability and population density in fish populations.
\newblock \emph{Nature}, 452, 344--347.

\bibitem[{Mosegaard \& Sambridge(2002)}]{Mosegaard2002}
\ifbool{MyRefNumbers}{\stepcounter{MyBibCount}\theMyBibCount.\\}{}Mosegaard, K.
  \& Sambridge, M. (2002).
\newblock Monte Carlo analysis of inverse problems.
\newblock \emph{Inverse Problems}, 18, R29.

\bibitem[{Okamoto \emph{et~al.}(2012)Okamoto, Schmitt, Holbrook \&
  Reed}]{Okamoto2012}
\ifbool{MyRefNumbers}{\stepcounter{MyBibCount}\theMyBibCount.\\}{}Okamoto,
  D.K., Schmitt, R.J., Holbrook, S.J. \& Reed, D.C. (2012).
\newblock Fluctuations in food supply drive recruitment variation in a marine
  fish.
\newblock \emph{Proceedings of the Royal Society B: Biological Sciences}, 279,
  4542--4550.

\bibitem[{Osenberg \emph{et~al.}(2002)Osenberg, St.~Mary, Schmitt, Holbrook,
  Chesson \& Byrne}]{Osenberg2002}
\ifbool{MyRefNumbers}{\stepcounter{MyBibCount}\theMyBibCount.\\}{}Osenberg,
  C.W., St.~Mary, C.M., Schmitt, R.J., Holbrook, S.J., Chesson, P. \& Byrne, B.
  (2002).
\newblock Rethinking ecological inference: density dependence in reef fishes.
\newblock \emph{Ecology Letters}, 5, 715--721.

\bibitem[{Pfister(1998)}]{Pfister1998}
\ifbool{MyRefNumbers}{\stepcounter{MyBibCount}\theMyBibCount.\\}{}Pfister, C.A.
  (1998).
\newblock Patterns of variance in stage-structured populations: evolutionary
  predictions and ecological implications.
\newblock \emph{Proceedings of the National Academy of Sciences}, 95, 213--218.

\bibitem[{Plummer(2013)}]{Plummer2013}
\ifbool{MyRefNumbers}{\stepcounter{MyBibCount}\theMyBibCount.\\}{}Plummer, M.
  (2013).
\newblock \emph{Just Another Gibbs Sampler (JAGS) Software}, 3rd edn.

\bibitem[{Quinn \& Deriso(1999)}]{Quinn1999}
\ifbool{MyRefNumbers}{\stepcounter{MyBibCount}\theMyBibCount.\\}{}Quinn, T.J.
  \& Deriso, R.B. (1999).
\newblock \emph{Quantitative fish dynamics}.
\newblock 1st edn.
\newblock Oxford University Press, New York.

\bibitem[{{R Core Team}(2014)}]{Rsoftware}
\ifbool{MyRefNumbers}{\stepcounter{MyBibCount}\theMyBibCount.\\}{}{R Core Team}
  (2014).
\newblock \emph{R: A language and environment for statistical computing}.
\newblock R Foundation for Statistical Computing, Vienna, Austria.

\bibitem[{Rideout \& Tomkiewicz(2011)}]{Rideout2011}
\ifbool{MyRefNumbers}{\stepcounter{MyBibCount}\theMyBibCount.\\}{}Rideout, R.M.
  \& Tomkiewicz, J. (2011).
\newblock Skipped spawning in fishes: More common than you might think.
\newblock \emph{Marine and Coastal Fisheries}, 3, 176--189.

\bibitem[{Rouyer \emph{et~al.}(2012)Rouyer, Sadykov, Ohlberger \&
  Stenseth}]{Rouyer2012}
\ifbool{MyRefNumbers}{\stepcounter{MyBibCount}\theMyBibCount.\\}{}Rouyer, T.,
  Sadykov, A., Ohlberger, J. \& Stenseth, N.C. (2012).
\newblock Does increasing mortality change the response of fish populations to
  environmental fluctuations?
\newblock \emph{Ecology Letters}, 15, 658--665.

\bibitem[{Schmitt \emph{et~al.}(1999)Schmitt, Holbrook \&
  Osenberg}]{Schmitt1999}
\ifbool{MyRefNumbers}{\stepcounter{MyBibCount}\theMyBibCount.\\}{}Schmitt,
  R.J., Holbrook, S. \& Osenberg, C. (1999).
\newblock Quantifying the effects of multiple processes on local abundance: a
  cohort approach for open populations.
\newblock \emph{Ecology Letters}, 2, 294--303.

\bibitem[{Schmitt \& Holbrook(1984)}]{Schmitt1984}
\ifbool{MyRefNumbers}{\stepcounter{MyBibCount}\theMyBibCount.\\}{}Schmitt, R.J.
  \& Holbrook, S.J. (1984).
\newblock Gape-limitation, foraging tactics and prey size selectivity of two
  microcarnivorous species of fish.
\newblock \emph{Oecologia}, 63, 6--12.

\bibitem[{Schmitt \& Holbrook(1986)}]{Schmitt1986}
\ifbool{MyRefNumbers}{\stepcounter{MyBibCount}\theMyBibCount.\\}{}Schmitt, R.J.
  \& Holbrook, S.J. (1986).
\newblock Seasonally fluctuating resources and temporal variability of
  interspecific competition.
\newblock \emph{Oecologia}, 69, 1--11.

\bibitem[{Shelton \& Mangel(2011)}]{Shelton2011}
\ifbool{MyRefNumbers}{\stepcounter{MyBibCount}\theMyBibCount.\\}{}Shelton, A.O.
  \& Mangel, M. (2011).
\newblock Fluctuations of fish populations and the magnifying effects of
  fishing.
\newblock \emph{Proceedings of the National Academy of Sciences}, 108,
  7075--7080.

\bibitem[{Shepherd(1982)}]{Shepherd1982}
\ifbool{MyRefNumbers}{\stepcounter{MyBibCount}\theMyBibCount.\\}{}Shepherd,
  J.G. (1982).
\newblock A versatile new stock-recruitment relationship for fisheries, and the
  construction of sustainable yield curves.
\newblock \emph{Journal du Conseil}, 40, 67--75.

\bibitem[{Winemiller \& Rose(1992)}]{Winemiller1992}
\ifbool{MyRefNumbers}{\stepcounter{MyBibCount}\theMyBibCount.\\}{}Winemiller,
  K.O. \& Rose, K.A. (1992).
\newblock Patterns of life-history diversification in North American fishes:
  implications for population regulation.
\newblock \emph{Canadian Journal of Fisheries and Aquatic Sciences}, 49,
  2196--2218.

\bibitem[{Worden \emph{et~al.}(2010)Worden, Botsford, Hastings \&
  Holland}]{Worden2010}
\ifbool{MyRefNumbers}{\stepcounter{MyBibCount}\theMyBibCount.\\}{}Worden, L.,
  Botsford, L.W., Hastings, A. \& Holland, M.D. (2010).
\newblock Frequency responses of age-structured populations: Pacific salmon as
  an example.
\newblock \emph{Theoretical Population Biology}, 78, 239--249.

\end{thebibliography}


\newcounter{MyBibCount}\providebool{MyRefNumbers}\begin{thebibliography}{1}
\expandafter\ifx\csname natexlab\endcsname\relax\def\natexlab#1{#1}\fi

\bibitem[{Okamoto \emph{et~al.}(2012)Okamoto, Schmitt, Holbrook \&
  Reed}]{Okamoto2012}
\ifbool{MyRefNumbers}{\stepcounter{MyBibCount}\theMyBibCount.\\}{}Okamoto,
  D.K., Schmitt, R.J., Holbrook, S.J. \& Reed, D.C. (2012).
\newblock Fluctuations in food supply drive recruitment variation in a marine
  fish.
\newblock \emph{Proceedings of the Royal Society B: Biological Sciences}, 279,
  4542--4550.

\end{thebibliography}
\end{document}


\pagenumbering{gobble} 
\title{Stochastic density effects on adult fish survival and implications for population fluctuations}
\author[1]{Daniel K. Okamoto\thanks{corresponding author: dokamoto@sfu.ca }\thanks{Present address: School of Resource and Environmental Management/Hakai Network\\ Simon Fraser University, Burnaby, BC V5A 1S6, Canada, Tel: (778) 686-9974
}}
\author[1,2]{Russell J. Schmitt\thanks{email: schmitt@lifesci.ucsb.edu}}
\author[1,2]{Sally J. Holbrook\thanks{email: sally.holbrook@lifesci.ucsb.edu}}
\affil[1]{Department of Ecology, Evolution and Marine Biology
 \\University of California Santa Barbara\\Santa Barbara, CA 93106 USA}
\affil[2]{Marine Science Institute\\ University of California Santa Barbara\\Santa Barbara, CA 93106 USA }
\renewcommand\Authands{ and }
\date{}
\newpage
\clearpage
\title{Appendices for ``Stochastic density effects on adult fish survival and implications for population fluctuations"}
\date{}

\maketitle

\newpage

\renewcommand{\thepage}{A\arabic{page}}
\setcounter{equation}{0}
\setcounter{page}{1}

\singlespacing
\section*{Appendix A: Supplemental methodological details for ``Stochastic density effects on adult fish survival and implications for population fluctuations"}
\newpage
\doublespacing

\renewcommand{\theequation}{A.\arabic{equation}}
\subsection*{Habitat and food density}
Estimates of food density for each site were calculated as  in \cite{Okamoto2012} and include the density of prey sampled from replicate 0.1 $\text{m}^2$ patches of foraging habitat in each year from which individual prey were counted and sized.  Food biomass was not normalized by the habitat percent cover on transects (as was done in \cite{Okamoto2012}) so habitat and biomass density of food could be evaluated independently.  Annual means of the biomass density of food averaged over all sites for each year of the time series were used as values for food availability in all analyses. 

\begingroup
\renewcommand{\section}[2]{}%
\bibliographystyle{ecol_let.bst}
\bibliography{Okamoto_et_al}
\clearpage
\endgroup 

\subsection*{Likelihoods and the posterior density}
The observation likelihood for eqn (2) is a truncated normal (to avoid negative adult density estimates) while the process likelihood for eqn (1)  is a lognormal.   Defining $\boldsymbol{\Theta}$ as the suite of parameters and other input variables, the normal observation likelihood for eqn (2)  is:
\begin{equation}
[\boldsymbol{\hat{A}}|\boldsymbol{A},\boldsymbol{\Theta}] \propto  \prod_{y=2}^Y\prod_{i=1}^N (\sigma_A)^{-1}\exp{\left[-\frac{(\hat{A}_{y,i} -{A}_{y,i})^2}{2 \sigma_A^2}\right]}, \ A_{y,i}\geq 0 \label{eq:3}
\end{equation}
The log-normal process likelihood for eqn (1) is:
\begin{equation}
[\boldsymbol{A}|\boldsymbol{\Theta}] \propto \prod_{y=2}^{Y} \prod_{i=1}^N (\sigma_\epsilon)^{-1}\exp{\left[-\frac{(\ln [A_{y,i}] -x_{y,i})^2}{2 \sigma_{\epsilon}^2}\right]}
\end{equation}
where 
\begin{equation}
x_{y,i}=
\begin{cases}
\ln[{A}_{y-1,i} {s}_{y-1,i}+{J}_{y-1,i}{s}_{y-1,i}g],&   \mathrm{if } \ y =2\\
\ln[{A}_{y-1,i} {s}_{y-1,i}+{J}_{y-1,i}{s}_{y-1,i}g]+\phi \epsilon_{y-1,i},&  \mathrm{if} \  y>2
\end{cases}
\end{equation}
and 
\begin{equation}
\epsilon_{y,i} = \ln[{A}_{y,i} {s}]-x_{y,i}
\end{equation}

The product of these likelihoods and the prior density yields the posterior density: 

\begin{equation}
 [\boldsymbol{A},\boldsymbol{\Theta}|\boldsymbol{\hat{A}}] \propto
[\boldsymbol{\hat{A}}|\boldsymbol{A},\boldsymbol{\Theta}][\boldsymbol{A}|\boldsymbol{\Theta}][\boldsymbol{\Theta}]\label{eq:5}
\end{equation}
\newpage
\subsection*{Added variable plots}
Added variable plots (also called partial regression plots) for Figure 2 were generated using the following algorithm:
\singlespacing
\begin{tabular}{p{6in}}
\toprule
I: for each iteration i = 1, 2, $\dots$, 1000: 
\begin{enumerate}
\item{From the markov chain of posterior parameter estimates $\boldsymbol{\Theta}$ extract a random iteration $\Theta_{j}$}
\item{Calcuate survival residuals adjusted to prey (y-axis in Figure 2A) by regressing survival estimates from $\Theta_j$ in each year (mean across sites) against food supply and calculating residuals}
\item{Calcuate survival residuals adjusted to adults (y-axis in Figure 2B) by regressing survival estimates in each year (mean across sites)from $\Theta_j$ against adult density estimates from in each year (mean across sites) from $\Theta_j$ and calculating residuals}
\item{Calcuate adult density residuals adjusted to prey x-axis in Figure 2A by regressing adult density estimates from in each year (mean across sites) from $\Theta_j$ against prey density and calculating residuals}
\item{Calcuate prey density residuals adjusted to adults x-axis in Figure 2B by regressing prey density against adult density estimates in each year (mean across sites) from $\theta_j$ and calculating residuals}
\end{enumerate}
II: Estimate mean and 95\% quantiles for each residual.\\
III: Plot appropriate means and quantile ranges. \\
IV: Overlay estimated mean slope from the posterior parameter distribution.\\
\bottomrule
\end{tabular}

\subsection*{Generalized effects of density dependent adult mortality on population variance}
\onehalfspacing
For each of the 100 replicate simulations within each of the 330,480 unique parameter combinations we generated a single 100 year time series of $\alpha = \bar{\alpha}e^{\boldsymbol{\epsilon}-\frac{\sigma^2}{2}}$.   $\boldsymbol{\epsilon}$ are autocorrelated errors with partial autocorrelation function $\phi$.   We achieved this by first calculating the variance ($\sigma^2$) of $\boldsymbol{\epsilon}$ required to generate the desired CV$[\alpha]$.  The asymptotic coefficient of variation of $\alpha = \bar{\alpha}e^{\boldsymbol{\epsilon}-\frac{\sigma^2}{2}}$ given that $\boldsymbol{\epsilon}$ is a gaussian process with variance $\sigma^2$ is: 

\begin{equation}
\text{CV}[\alpha]=\frac{\sqrt{\text{var}[\alpha]}}{\bar{\alpha}}=e^{\sigma^2}\!\!-1
\end{equation}
Thus, given the desired CV$[\alpha$] we calculate the correspoding $\sigma$ as:

\begin{equation}
\sqrt{\log{(\text{CV}[a])^2+1)}}=\sigma
\end{equation}
To generate $\boldsymbol{\epsilon}$, we simulate 100 random innovations ($\boldsymbol{e} \in e_1, e_2, \dots, e_{100}$) from a standard normal. We then convert the time series $\boldsymbol{e}$ to an AR(1) process with the desired properties.  
  For the variance of $\epsilon$ to be $\sigma^2$ the variance of the innovations $\boldsymbol{e} $ must be $\sigma^2({1-\phi^2})$ (the long-term variance of a positively correlated time series is greater than the variance of the innovations or a short subset of the series). Thus, we generate $\boldsymbol{\epsilon}$ as follows:
  
\begin{align}
\epsilon_{y+1}&= 
\begin{cases}
e_y\sigma\sqrt{1-\phi^2},&   \mathrm{if } \ y =1\\
\phi\epsilon_y+e_y\sigma\sqrt{1-\phi^2},&   \mathrm{if } \ y >1\\
\end{cases}\\
\boldsymbol{e}&\sim \text{normal}(0,1)
\end{align}

Given $\bar{\alpha}, \beta$, and $\gamma$ as well as newly generated $\boldsymbol{\epsilon}$ we numerically solved for the value of density-independent mortality $z$ that produced the desired long-term mean survival rate ($\bar{S}$).  We combined this estimated  $z$ with $\boldsymbol{\epsilon}, \bar{\alpha}, \beta, $ and $\gamma$, produced final simulation results and averaged the temporal CV in adult abundance over the 100 replicate 100 year time series.  

\newpage

\pagenumbering{arabic}
\renewcommand{\thepage}{B\arabic{page}}
\singlespacing
\section*{Appendix B: Supplemental tables and figures for ``Stochastic density effects on adult fish survival and implications for population fluctuations"}
\setcounter{table}{0}
\renewcommand{\thetable}{B.\arabic{table}}
\setcounter{figure}{0}
\renewcommand{\thefigure}{B.\arabic{figure}}

\begin{landscape}
\begin{table}[ht]
\centering
\caption[Parameter description and prior specification]{Parameter description for the Shepherd Model under the SSVS procedure along with prior specification.   The Logistic Model uses the same specification with the exception that there is no scale coefficient.  }
 \begin{threeparttable}
\begin{tabular}{cll}
\toprule
Parameter& Description& Prior \\
\midrule
$\text{logit}\left({e^{-{z}_{i}}}\right)$& logit-scale density independent annual survival rate for the $i$th site& $\text{N}\left(\mu[\text{logit}\left(e^{-z}\right)],  \sigma[\text{logit}\left(e^{-z}\right)]\right)$ \\
$\mu[\text{logit}\left(e^{-z}\right)]$& mean of $\text{logit}\left({e^{-{z}_{i}}}\right)$ & $\text{N}\left(0,  1.5\right)$ \\
$\sigma[\text{logit}\left(e^{-z}\right)]$& among site variance in $\text{logit}\left({e^{-{z}_{i}}}\right)$& $\text{Uniform}\left(0,  1.5\right)$  \\
\\
${{\beta}_{0}}_{i}$& scale coefficient for the $i$th site& $\text{N}\left(\mu[{\beta}_{0}],  \sigma[{\beta}_{0}]\right)$\\
$\mu[{\beta}_{0}]$& mean scale coefficient& $\text{Uniform}\left(-5,  5\right)$\\
$\sigma[{\beta}_{0}]$& among site variance in scale coefficient  & $\text{Uniform}\left(0,  5\right)$ \\
\\
${\gamma}$& density dependent shape coefficient\mtnote{$\dagger$}& $\left(1-{\psi}_{\gamma}\right)\text{N}\left(0,0.05\right)+{\psi}_{\gamma}\text{N}\left(0,5\right)$\\
${\beta}_{2},\dotsc,{\beta}_{6}$& other variable coefficients\mtnote{$\dagger$}& $\left(1-{\psi}_{i}\right)\text{N}\left(0,0.05\right)+{\psi}_{i}\text{N}\left(0,5\right)$\\
$g$& annual juvenile maturation rate&$\text{Uniform}\left(0, 1\right)$ \\
$\phi$& first-order autoregressive coefficient& $\text{Uniform}\left(0.9, 0.9\right)$ \\
${\sigma}_{{\epsilon}_{i}}$& process error for the $i$th site& $\text{Uniform}\left(0, 100\right)$ \\
\bottomrule
\end{tabular}
\begin{tablenotes}
\footnotesize
            \item[$\dagger$] When models were refit using only the covariates with strong support, those priors were replaced with uniform density from -10 to 10 (the realistic range of support).
        \end{tablenotes}  
 \end{threeparttable}
 \label{table:priors}
 \end{table}
\end{landscape}
\newpage

\begin{figure}[ht]
\begin{center}
\includegraphics{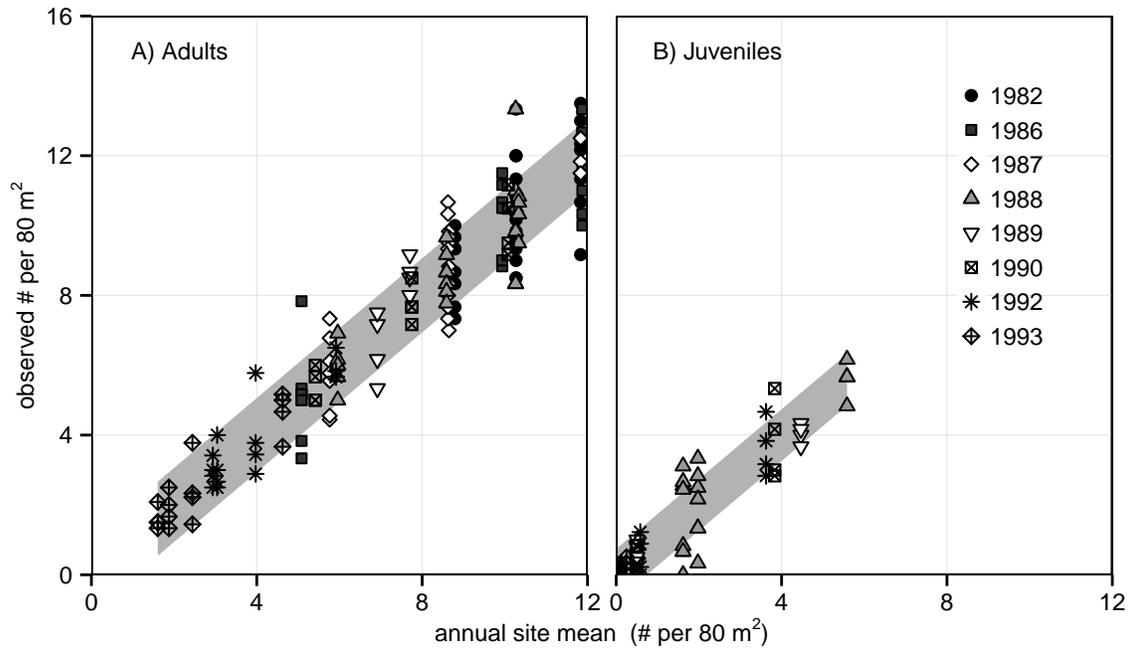}
\end{center}
\caption[Training dataset used to estimate observer error]{Training dataset illustrating observed site densities versus the empirical mean for that site within that year.  Fish surveys were replicated repeatedly over the late summer and fall season.   The grey bar represents the estimated standard error about the 1:1 line ($\sigma_{A}=1.06$  for adults [A] and $\sigma_{J}= 0.74$ for juveniles [B]).   The standard error was estimated independently for adults and juveniles using generalized least squares.  There was no significant evidence that during this period within site error variance increased as a function of the mean, among years or among sites.   These data represent sites that were visited by divers during late summer and early fall 4-16 times per year for replicate observation.  Juveniles were not counted prior to 1988. }  
\label{figure:F2A1}
\end{figure}
\newpage

\begin{figure}[ht]
\begin{center}
\includegraphics{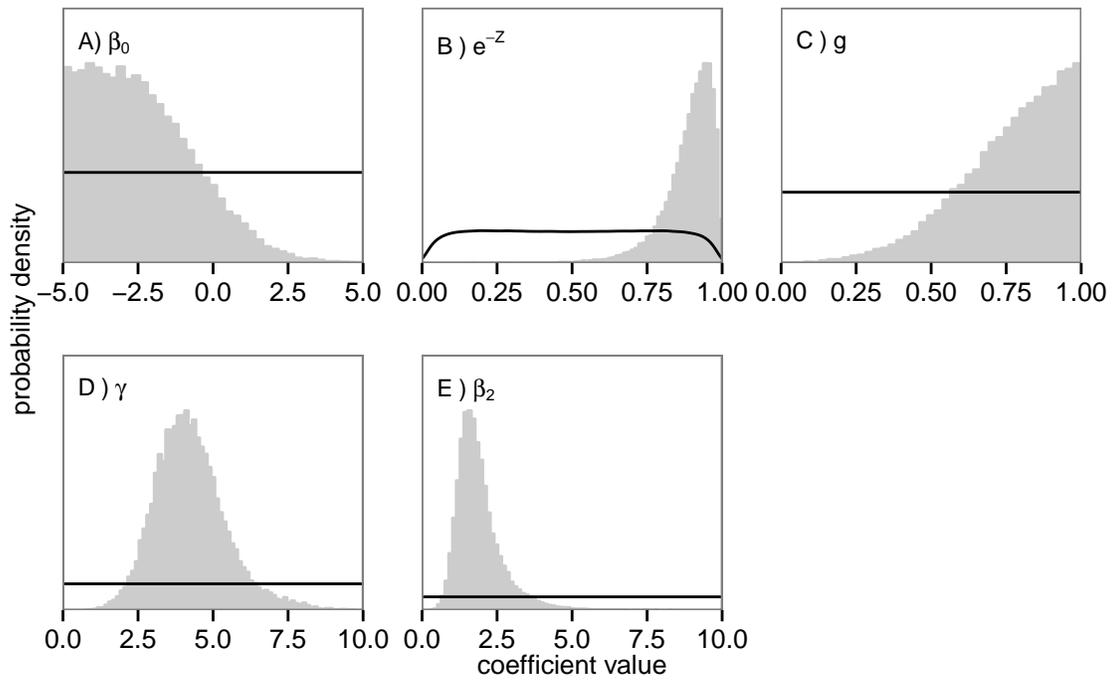}
\end{center}
\caption[Posteriors versus priors]{Posterior (grey histograms) and prior (black line) densities for primary parameters in the Shepherd Model.  Prior probabilities and parameter descriptions are shown in Table \ref{table:priors}.}
\label{figure:F2A2}
\end{figure}
\newpage

\begin{figure}[ht]
\begin{center}
\includegraphics{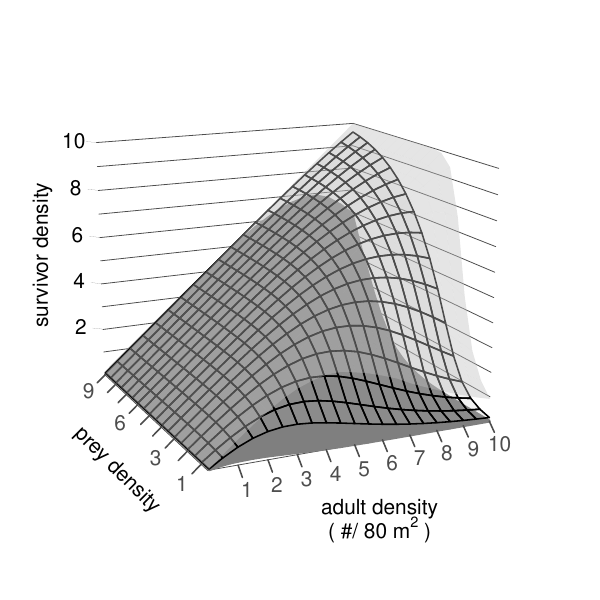}
\end{center}
\caption[Survivorship response surface under the Logistic model]{Survivorship as predicted for a given adult density and the observed range of prey density using posterior samples from the Logistic Model [eqn (4)].  The gridded surface represents the posterior mean expectation, while the grey surfaces represent the 95\% posterior credible set for survival (z-axis) in each x-y combination.}  
\label{figure:F2A3}
\end{figure}
\newpage

\begin{figure}[!ht]
\begin{center}
\includegraphics[width= 6in]{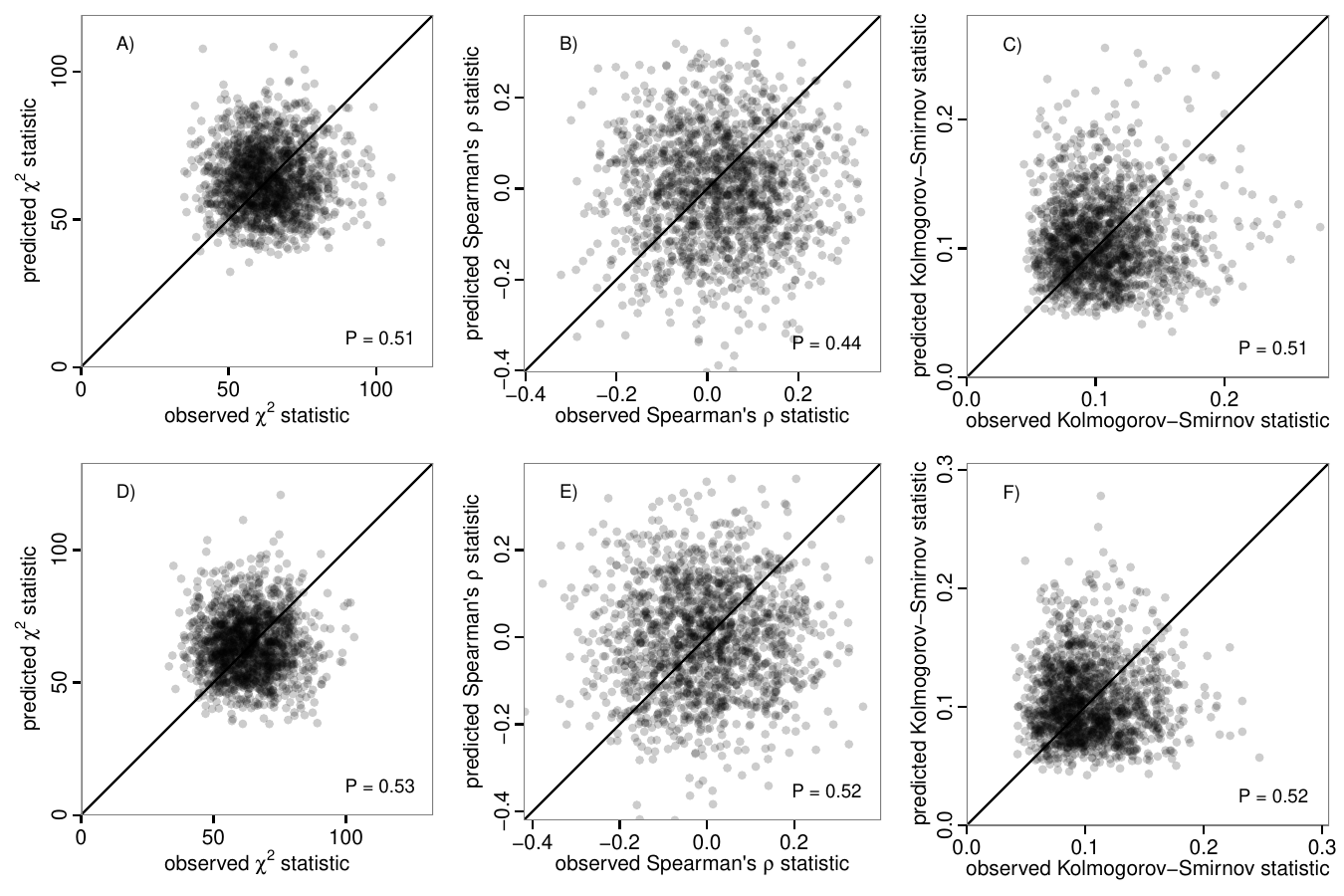}
\end{center}
\caption[Posterior predictive checks]{Posterior predictive checks of model assumptions and fit for the Shepherd model (A-C) and the logistic model (D-F).  Metrics include $\chi^2$ goodness of fit-test statistics measuring model inadequacy (A, D), Spearman rank-order correlation statistics testing for correlation between residuals and the response values (B, E) and Kolmogorov-Smirnov test statistics comparing adequacy of the lognormal process likelihood function (C, F).  Metrics indicate problems for the model if the point cloud lies substantially below the $45\degree$  1:1 line which leads to lower p-values.   P-values indicate the probability that the observed value is as or less extreme than under the posterior predictive distribution.  Y-axes represent the predicted value, while x-axes represent realized statistics.}  
\label{figure:F2A4}
\end{figure}
\newpage